\documentclass[12pt]{iopart}
\usepackage{graphicx}
\usepackage{wrapfig}
\begin{document}

\title{Strangeness and charm in QCD matter}

\author{B K\"ampfer\footnote[3]{kaempfer@fz-rossendorf.de.
Supported by BMBF 06DR121, GSI, and EU-I3-HP.} 
and M Bluhm}

\address{Forschungszentrum Rossendorf/Dresden, PF 510119, D-01314 Dresden, Germany}

\begin{abstract}
Strangeness and charm degrees of freedom in strongly interacting matter
are discussed within a quasi-particle model adjusted to
lattice QCD data. The model allows to extrapolate lattice QCD data 
to large baryo-chemical potential. We outline the
thermal evolution of matter in the early universe at and slightly after confinement
and comment briefly on charm dynamics in relativistic heavy-ion collisions.
\end{abstract}

\section{Introduction}

The recent advent of numerical evaluations of the thermodynamics of strongly interacting
matter, based on first principles, delivers the equation of state below and above
the deconfinement temperature $T_c$.
Equipped with this knowledge the equation of state can be extrapolated in a large
region of interest. We perform here such an extrapolation employing our quasi-particle
model. Having the equation of state at our disposal we consider the strangeness
and charm excitations (Section 2). Then we follow the evolution of strongly
interacting matter in the universe during confinement (Section 3).
We address also the strangeness fraction of matter after cosmic confinement  
and the hadron freeze-out (Section 4). 
Finally we comment briefly on charm dynamics 
in relativistic heavy-ion collisions (Section 5).     

\section{Strangeness and charm in thermalized matter} \label{QPM}

\begin{figure}[t]
~\center
\includegraphics[width=0.399\textwidth]{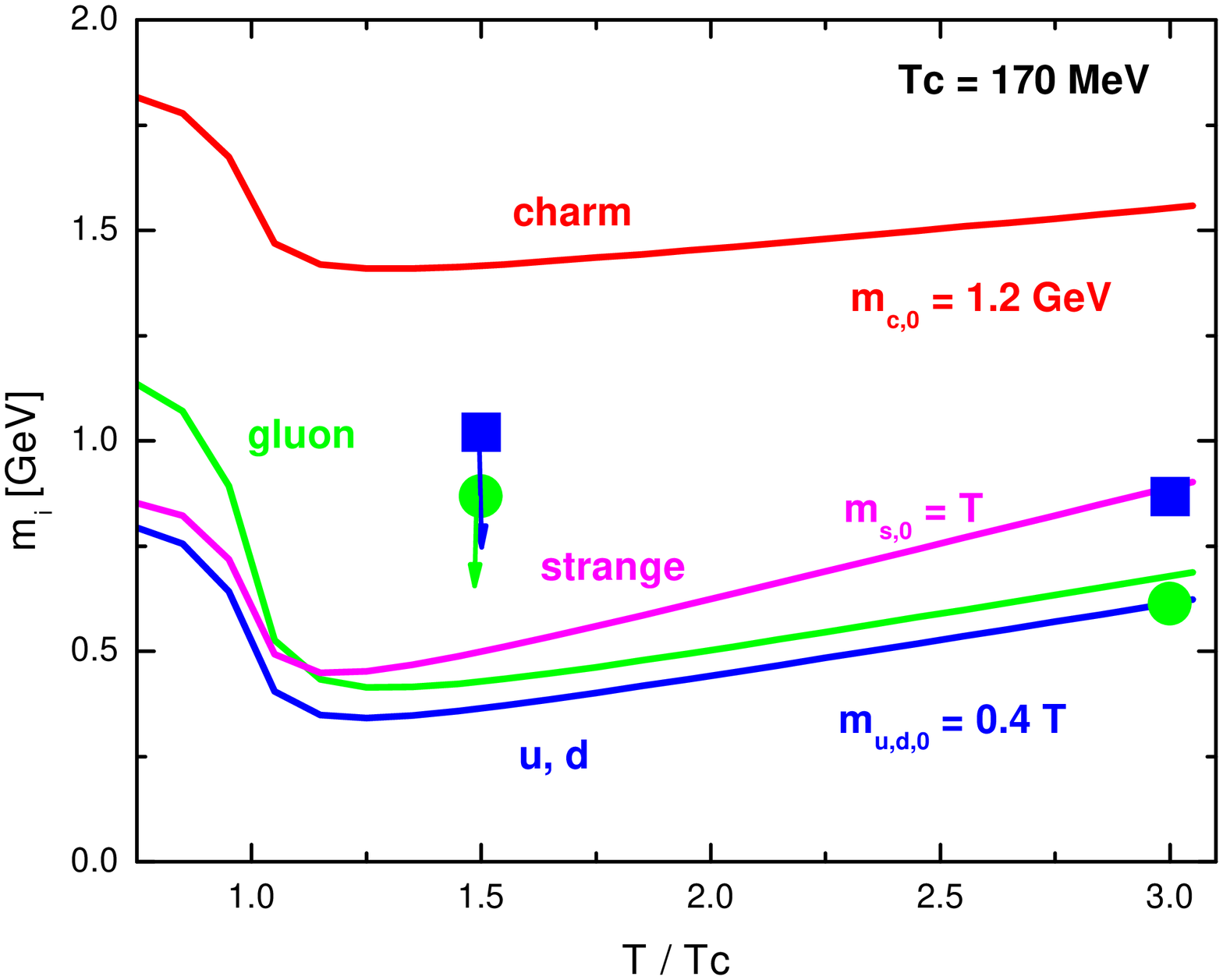} \hspace*{12mm}
\includegraphics[width=0.39\textwidth]{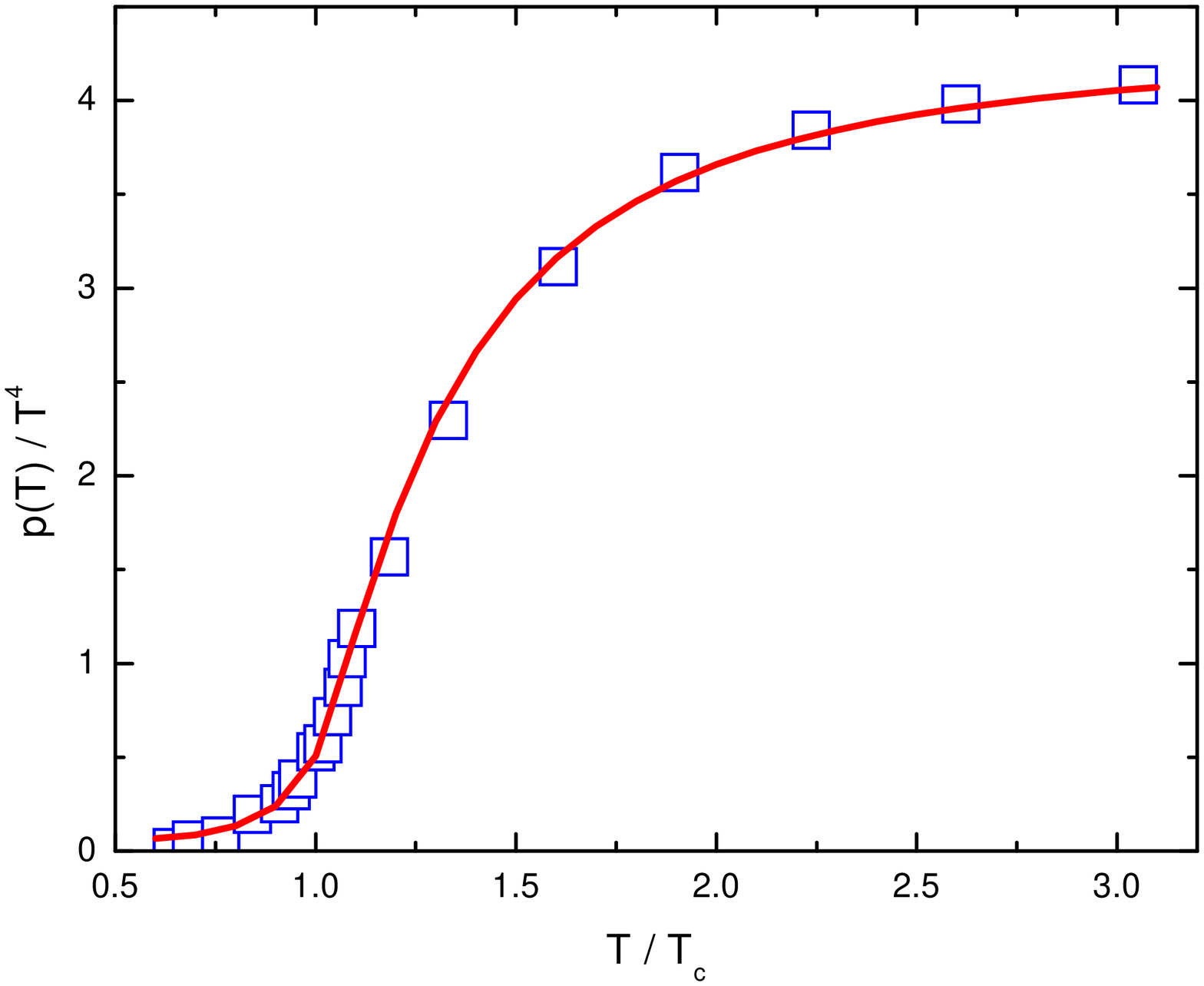}
\caption{Energies $m_i$ of quasi-particle excitations (left panel,
quasi-particle energies from \cite{Petretsky}
are depicted by symbols) adjusted to 
lattice QCD data \cite{Nf2+1} for the scaled pressure $p/T^4$ 
at vanishing chemical potential (right panel).}
\label{fig_EoS}
\end{figure}

Our model \cite{Peshier,Bluhm} rests on the idea
that strongly correlated systems can be described
in terms of quasi-particles. In Fig.~\ref{fig_EoS} the quasi-particle energies at vanishing
momentum are shown when adjusting the model to the lattice QCD data for 
2 + 1 flavors \cite{Nf2+1}.
The quasi-particle energies are of the order
of $0.5 \cdots 1$ GeV. Strange excitations are similar to excitations
of gluons and $u, d$ quarks. If the excitations are such massive one could
argue that also charm is copiously excited. However, the interaction energy
and rest mass of charm add up to such large values that charm is thermally
suppressed, as expected. 

In Fig.~\ref{fig_EoS} also quasi-particle energies directly derived
from lattice QCD evaluations \cite{Petretsky} are depicted.
These quasi-particle measurements \cite{Petretsky}
are performed for quite a different lattice configuration. The comparably large
values at $1.5 T_c$ (which however represent an upper limit as indicated) 
have been considered in \cite{Shuryak} as hint to additional degrees  
of freedom contributing to the pressure. Further dedicated measurements on the 
same lattices, where also the equation of state is evaluated,
and in the proper momentum range $k \sim T$, which is relevant for bulk thermodynamics,
are needed to clarify that issue.

Our model is successfully applicable to describe the thermodynamics of the
quark-gluon fluid at non-vanishing chemical potential \cite{Bluhm}.
One can then extrapolate the equation of state into a large
interval of chemical potentials not yet accessible to lattice evaluations.
This knowledge is of importance for several estimates with respect
to the CBM project at FAIR.
Results will be reported elsewhere.    

\section{Cosmic confinement dynamics}

\begin{figure}[t]
~\center
\includegraphics[width=0.30\textwidth]{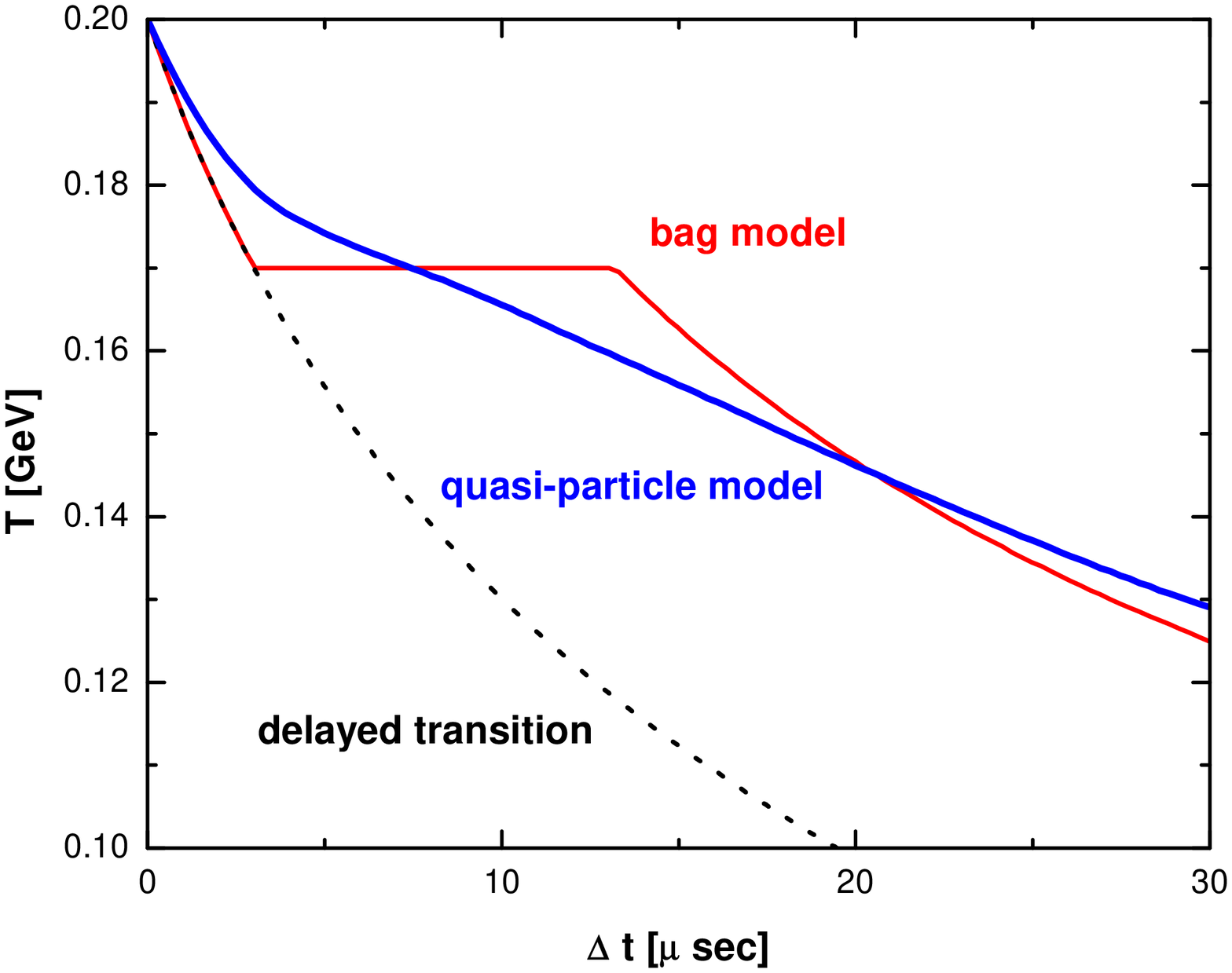} \hfill
\includegraphics[width=0.30\textwidth]{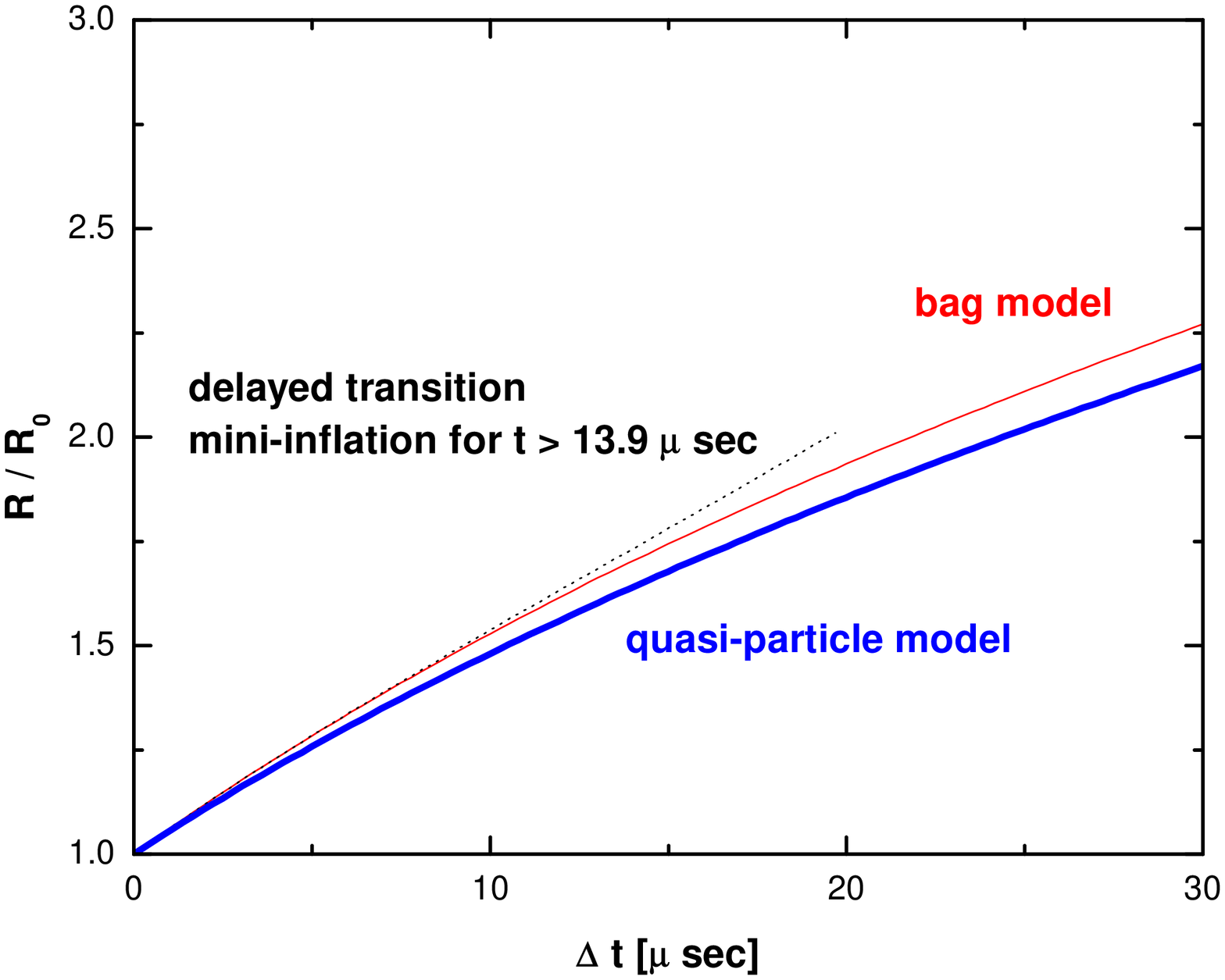} \hfill
\includegraphics[width=0.30\textwidth]{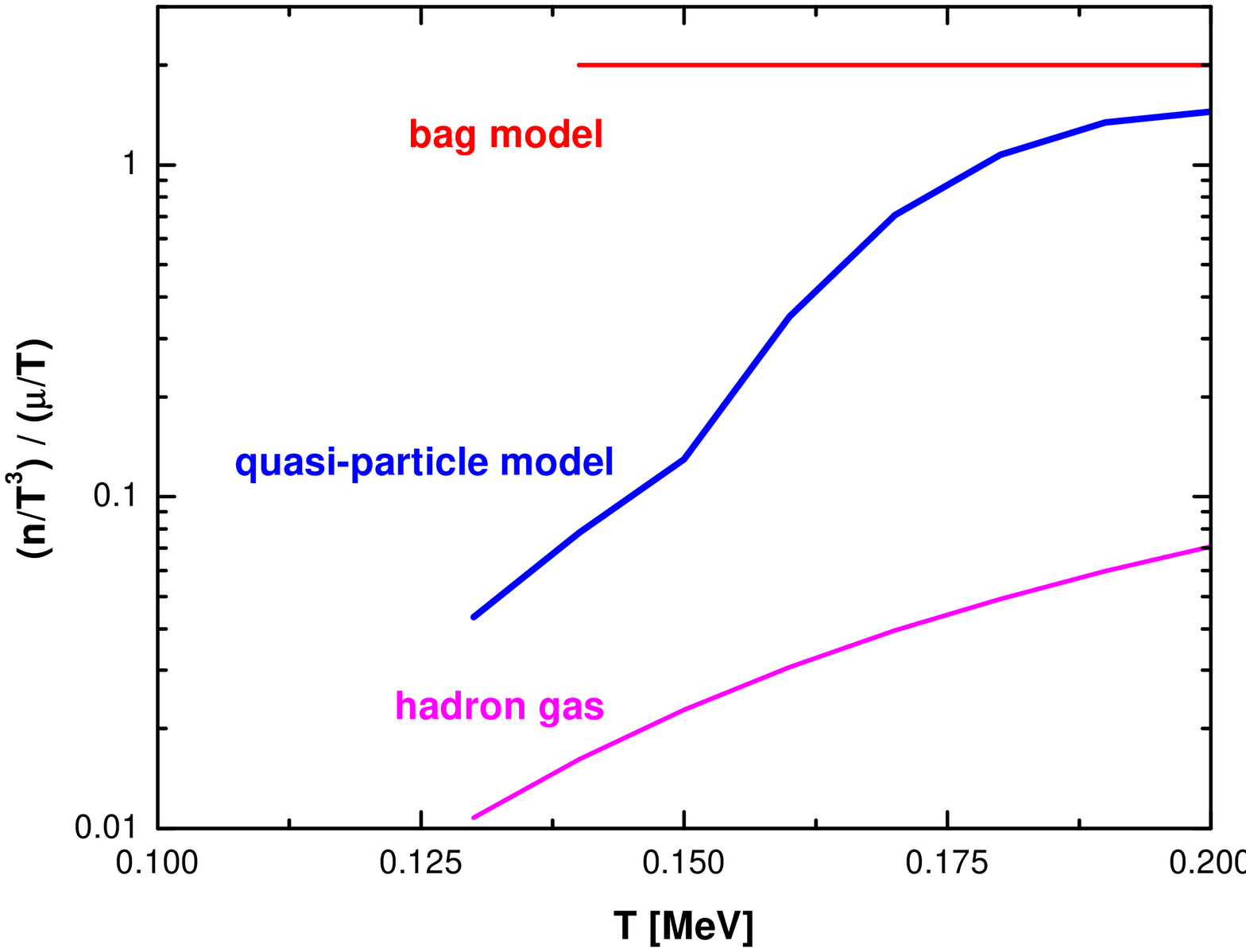}
\caption{Temperature $T$ (left panel) and scale factor $R$ (middle
panel) as a function of the time
(with zero point arbitrarily adjusted at $T = 200$ MeV).
The right panel exhibits the scaled baryon density $n/T^3 (\mu_q/T)^{-1}$
as a function of the temperature.}
\label{fig_confinement}
\end{figure}

The evolution of matter in an isotropic, homogeneous and flat universe
is determined by the Friedmann equations and baryon conservation
\begin{equation}
\dot R = {\cal C} R \sqrt{e}, \quad 
\dot e = - 3 {\cal C} (e + p) \sqrt{e}, \quad
(n_B R^3)\dot{} = 0,
\end{equation} 
where ${\cal C} = \sqrt{8 \pi /3} / M_{Pl}$ with
$M_{Pl}$ as Planck mass, $R$ stands for the scale factor,
and $p$, $e$ and $n_B$ denote the pressure, energy density
and baryon density, respectively.
In the early universe $\mu / T \ll 1$ due to
$\eta \equiv (n_\gamma / n_B)_0 \approx 10^{10}$.
Being aware of the need of a systematical chiral extrapolation, we tentatively
employ the parametrization described in Section \ref{QPM} 
and extend to finite chemical potential $\mu$ \cite{Bluhm}
with $T_c = 170$ MeV. Adding the background of leptons and photons
we arrive at the results displayed in Fig.~\ref{fig_confinement}.
In contrast to the bag model, the temperature is continuously dropping
at confinement (left panel). The differences of the time evolution of the 
scale factor are tiny for various scenarios (middle panel). In \cite{Jenkowsky}  
it was argued that a strong supercooling with dominating vacuum energy,
represented by the bag constant, could cause a mini-inflationary era.
With the present equation of state such a scenario is unlikely.

The bag model equation of state would deliver a large
difference of baryon densities under equilibrium condition 
(equal temperatures, chemical potentials and pressures)
in the confined and deconfined phases (right panel). It was thought that
in a departure from equilibrium at the end of the confinement transition
the baryon charge is concentrated in the last islands of the deconfined matter
which afterwards transforms to energetically more favorable strange
matter \cite{Witten}. The present equation of state from lattice QCD, however, 
disfavors such a scenario.
 
\section{Cosmic hadron freeze-out}

\begin{figure}[t]
~\center
\includegraphics[width=0.36\textwidth]{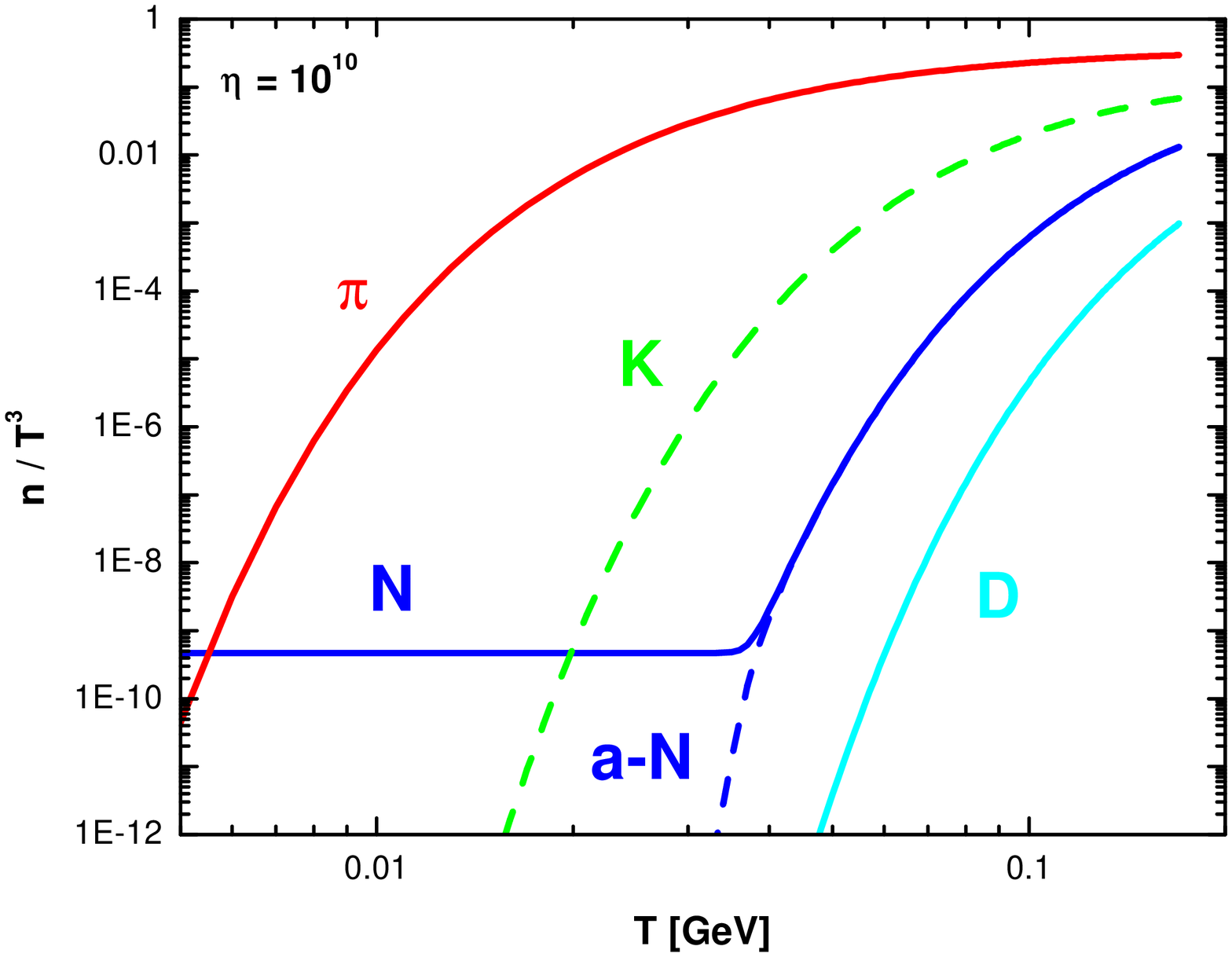} \hspace*{12mm}
\includegraphics[width=0.36\textwidth]{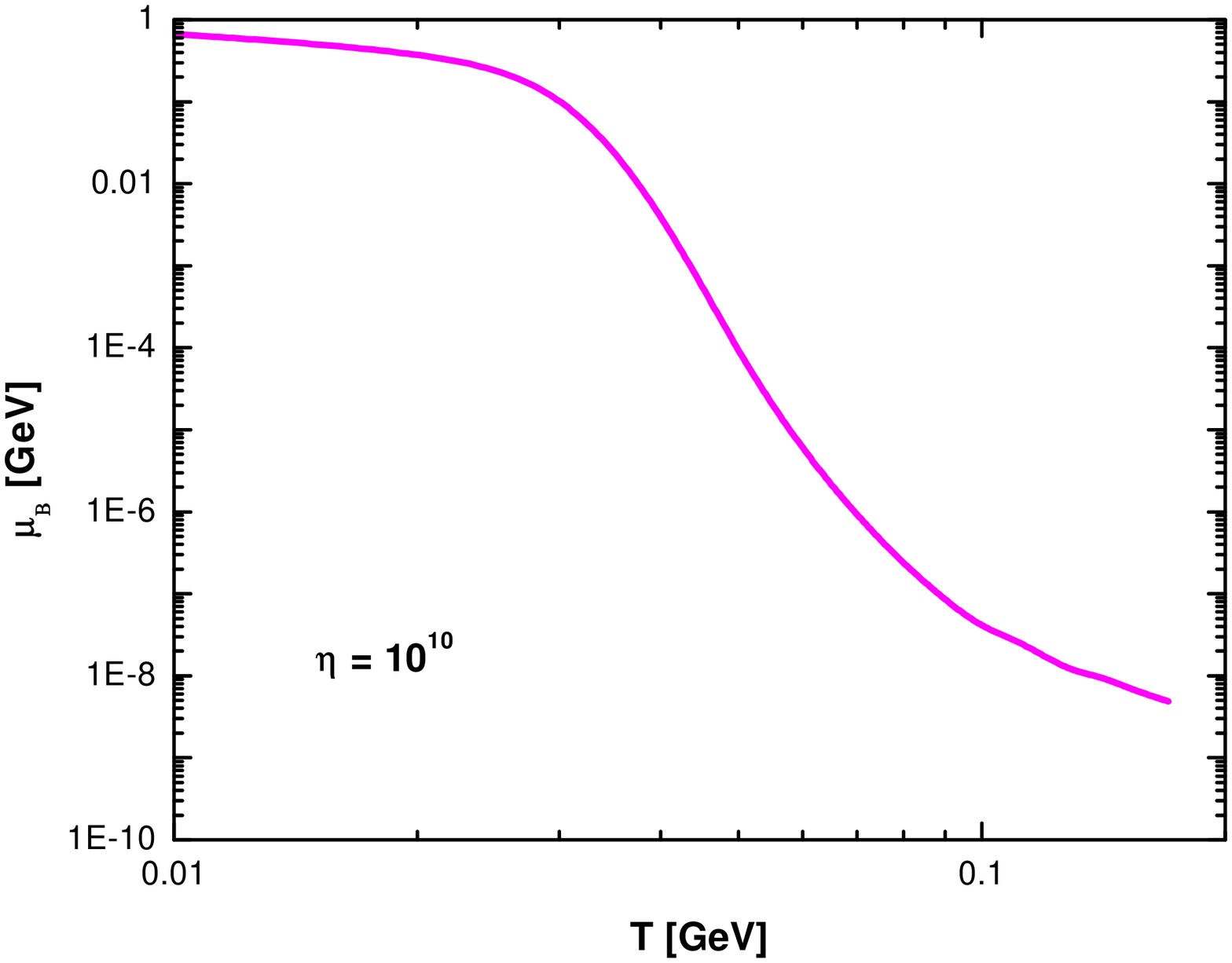}\\[7mm]
\includegraphics[width=0.36\textwidth]{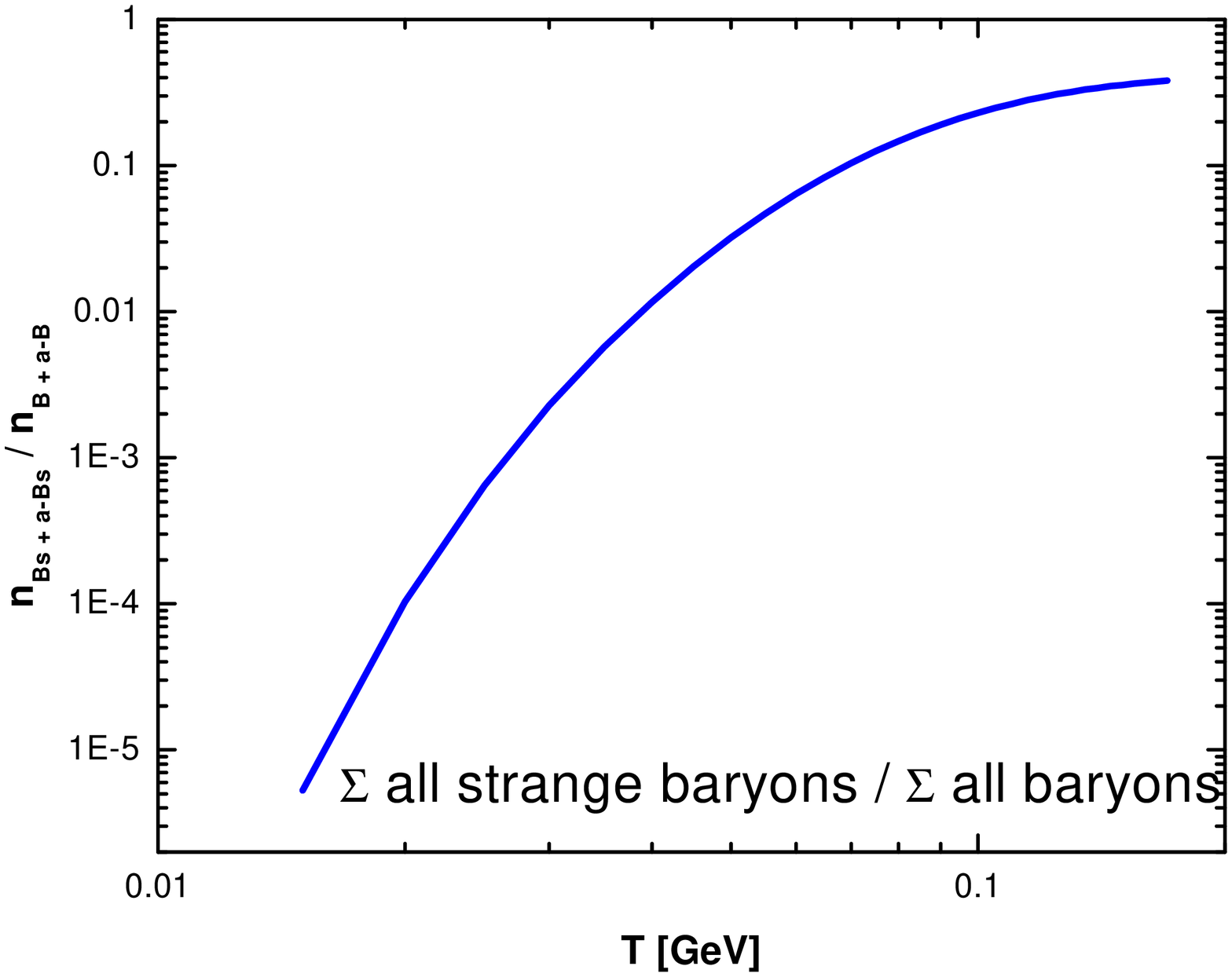} \hspace*{12mm}
\includegraphics[width=0.36\textwidth]{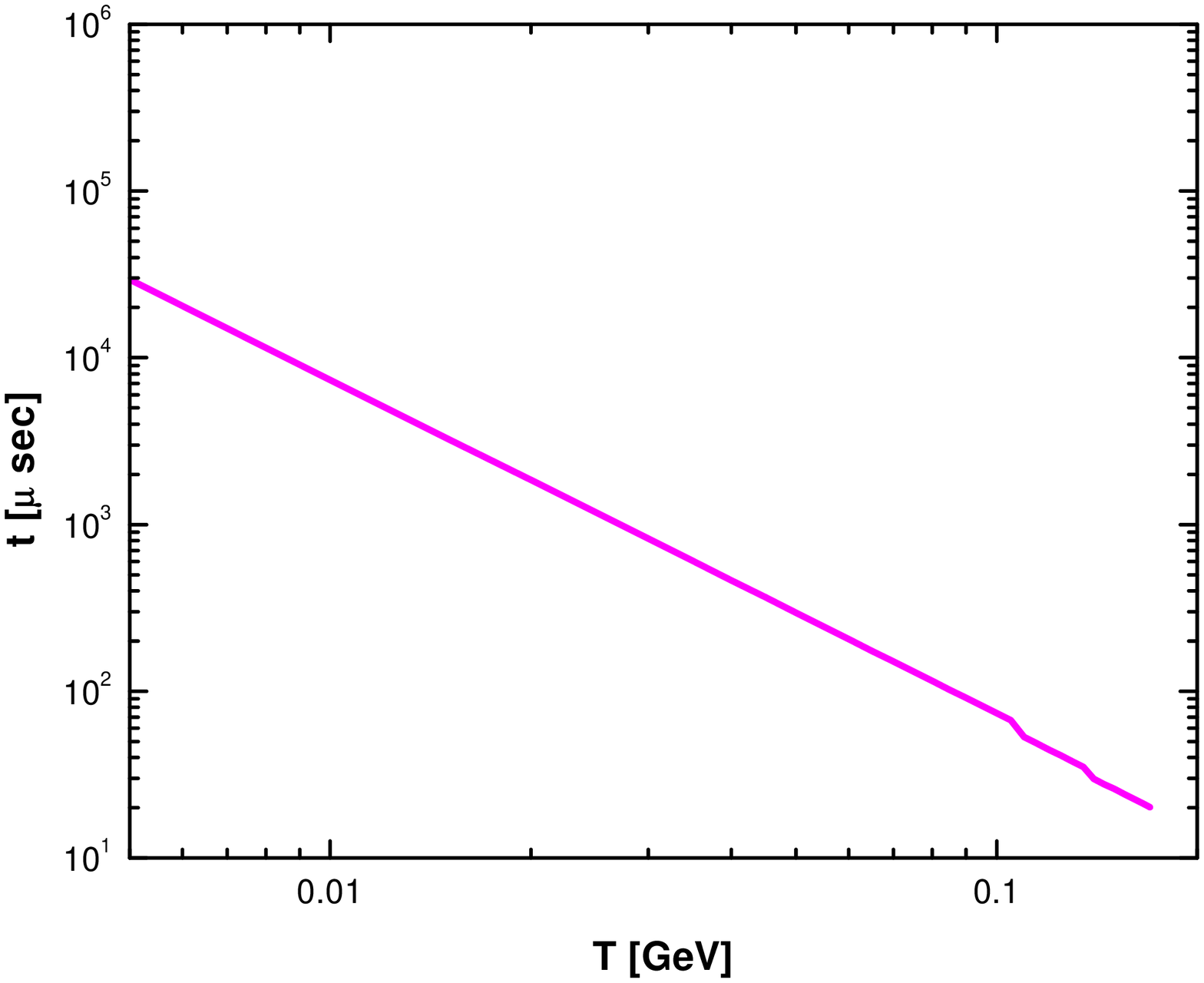}
\caption{Temperature dependence of 
scaled densities of selected hadrons (upper left panel),
baryo-chemical potential (upper right panel),
ratio of strange baryons to all baryons (lower left panel), and
elapsing time (lower right panel).}
\label{fig_hadron_freeze_out}
\end{figure}

The evolution of the strongly interacting matter
component in the early universe can further be followed
by employing the resonance gas model.
Note that the total energy density and pressure below $T_c$ 
are dominated by leptons and photons.
In Fig.~\ref{fig_hadron_freeze_out} various quantities of interest are exhibited.
In the upper left panel the densities of a few selected hadrons are
depicted. Pions dominate some time, kaons are subdominant,
open charm is completely negligible. The tiny baryon excess,
described by the present photon-to-baryon ratio $\eta \approx 10^{10}$
is unimportant down to temperatures ${\cal O} (40$ MeV), i.e., the densities
of baryons ($N$) and anti-baryons ($a-N$) are nearly the same. Below 40 MeV, however,
baryon conservation drives the chemical potential towards 1 GeV
(upper right panel) causing a constant comoving baryon density, while
the anti-baryon density drops rapidly. This explains the disappearance
of anti-matter. Briefly after confinement the baryon component was quite
strange: About 30\% of the baryons, constituting now the hard core of
visible matter in the universe, were in form of $\Lambda$s (lower left panel). 

It is often claimed that matter is produced in relativistic heavy-ion 
experiments under similar circumstances as in the early universe.  
While at RHIC, at chemical
freeze-out, a baryo-chemical potential ${\cal O} (10$ MeV) is found
\cite{Cleymans}, in the early universe, at comparable temperatures, the
chemical potential is ${\cal O} (1$ eV) (upper right panel, cf.\ also \cite{Rafelski}). 
Besides, there is a factor $10^{18}$ difference in time scales 
(lower right panel). 

\section{Charm dynamics at RHIC}

Our finding in Section \ref{QPM} suggests that charm 
is not noticeably created by thermal excitations
at physically relevant temperatures.
Rather, charm is created by hard processes in the initial stage
of heavy-ion collisions. One interesting question
concerns the energy loss of charm quarks. Due to the semileptonic
decays of $D$ mesons the momentum distribution
of inclusive decay electrons should be modified by the
energy loss. Our first analysis \cite{Gallmeister} did not evidence
a signal for such an energy loss. It could be that the
dead cone effect \cite{Ronny}, the Ter-Mikaelian effect \cite{Pavlenko}
and the Landau-Pomeranchuk effect (cf.~\cite{Gyulassy}) 
suppress the energy loss. As shown in 
\cite{Gallmeister}, the di-electrons from correlated
semileptonic decays of open charm are more sensitive to energy loss effects.
The recent run-4 results at RHIC will shed further light on that issue.

\end{document}